# Evidence for a radiation belt around a brown dwarf

J. B. Climent[1,2]\*, J. C. Guirado[1,3], M. Pérez-Torres[4,5], J. M. Marcaide[6,7], L. Peña-Moñino[4]

[1]Departament d'Astronomia i Astrofísica, Universitat de València; Burjassot, E-46100, Spain.

[2]Universidad Internacional de Valencia; Valencia, E-46002, Spain.

[3]Observatori Astronòmic, Universitat de València, Parc Científic; Paterna, E-46980, Spain.

[4]Instituto de Astrofísica de Andalucía, Consejo Superior de Investigaciones Científicas; Granada, E-18008, Spain.

[5]Facultad de Ciencias, Universidad de Zaragoza; Zaragoza, E-50009, Spain.

[6]Real Academia de Ciencias Exactas, Físicas y Naturales de España; Madrid, E-28004, Spain.

[7]Donostia International Physics Center; San Sebastián, E-20018, Spain.

\*Corresponding author. Email: j.bautista.climent@uv.es

**Ultracool dwarfs are a category of astronomical objects including brown dwarfs and very low-mass stars. Radio observations of ultracool dwarfs have measured their brightness as a function of time (light-curves) and spectral energy distributions, providing insight into their magnetic fields. We present spatially resolved radio observations of the brown dwarf LSR J1835+3259 using very long baseline interferometry, which show extended radio emission. The detected morphology is consistent with the presence of a radiation belt. Comparison with models indicates the radiation belt contains energetic particles confined by magnetic mirroring. Similar to the Jupiter case, we argue that radio emitting ultracool dwarfs possess dipole-ordered magnetic fields with radiation belt-like morphologies and aurorae.**

**Main Text:**

Ultracool dwarfs (UCDs) are low-mass stars and sub-stellar objects, classified as having spectral types later than M6. Only a small fraction of UCDs have detectable radio emission at GHz frequencies (*1*). In those that do, the radio emission can be separated into i) a circularly polarized component which appears as bursts, modulated by the object's rotational period, and ii) a quiescent slowly-varying emission component, with a low degree of circular polarization. The bursting component has been linked to the electron cyclotron maser instability (ECMI, *2*) mechanism, a coherent radio emission process responsible for auroral activity on planets (*3, 4*). However the details of the emission mechanism could vary between UCDs and planets (*5, 6*). The quiescent component is usually interpreted as due to radio emission from mildly or ultra relativistic electrons spiraling in magnetic fields [gyrosynchrotron and synchrotron mechanisms, respectively (*7, 8*)] originating in the UCD's corona and/or radiation belts (*9-11*). Radio emission in UCDs has mostly relied upon the analysis of their light curves (brightness as a function of time), spectral energy distributions, and dynamic spectra. Attempts to spatially resolve the radio emission using very long baseline interferometry (VLBI) have only detected a few objects (*12-14*) and found their radio images were unresolved.

**Radio observations of LSR J1835+3259**

On 2021 June 5, we observed the M8.5 brown dwarf  LSR J1835+3259 (also cataloged as 2MASS J18353790+3259545) using the European VLBI Network (EVN). LSR J1835+3259 exhibits a rapid rotation [rotational period $2.84140 \pm 0.00039$ hr; (*15*)], and is located at a distance of $5.6885 \pm 0.0015$ parsecs [pc; (*16*)]. It was already known to have strong radio emission, in both the bursting and quiescent components (*3, 9, 17-19*). Our observations resolve a complex radio morphology for this UCD (Fig. 1). The radio emission extends up to 3 milliarcseconds (mas), which corresponds to 0.017 astronomical units (au) at our adopted distance, or 33 stellar radii ($R_\star$), which is $1.07 \pm 0.05$ Jupiter radii ($R_{Jup}$) for LSR J1835+3259 (*20*).

The light curve of our observation (Fig. 2) contains two bursts of left circularly polarized (LCP) radio emission at rotation phases $\phi_1 = 0.38 \pm 0.02$ and $\phi_2 = 1.33 \pm 0.02$, with LCP flux densities of $3.7 \pm 0.2$ millijansky (mJy) and $1.5 \pm 0.2$ mJy, respectively. We interpret these as ECMI emission and refer to these two bursts as B1 and B2 hereafter. The time interval between the bursts' maxima is $2.70 \pm 0.02$ hr, which is close to (but different from) the rotation period ($2.84140 \pm 0.00039$ hr, *15*). The right circularly polarized (RCP) emission varies slowly during the observation, reaching a maximum before the bursts occur. Both LCP and RCP quiescent flux densities are very similar, as expected for a weakly polarized, non-bursting component of the LSR J1835+3259 radio emission.

We reconstructed images of the radio emission *(21)* of LSR J1835+3259 during burst B1 (rotation phase interval: $0.29 < \phi < 0.47$; Fig. 3A). These show that the non-bursting component (unpolarized, traced by the total intensity of radiation emitted or Stokes I emission) is spatially resolved into two radio sources separated by $2.3 \pm 0.2$ mas in the east-west direction, which are largely unpolarized (circular polarization <10%). In contrast, the bursting ECMI emission component (traced by circularly polarized emission or Stokes V emission) appears approximately halfway between the quiescent components. A similar morphology occurs during burst B2 ($1.35 < \phi < 1.52$; Fig. 3B), although the separation of the double structure is $1.2 \pm 0.2$ mas, approximately half that measured during the first burst.

## Interpretation of the radio images

Optical astrometry of LSR J1835+3259 (*21*) gives a location coinciding with the ECMI emission (within uncertainties, Fig. 3). ECMI emission is thought to be produced near the poles of a UCD, in the walls of an emission cone oriented almost perpendicular to the magnetic field lines (*22, 23*). This implies that the magnetic axis of LSR J1835+3259 is nearly perpendicular to the line of sight, at least at the times of Figs. 3A and B. We consider two possible interpretations for the quiescent double structure: i) bipolar non-thermal emission (a magnetic axis with position angle P.A. ~ 90°), similar to that observed in other low-mass stars [such as UV Ceti (*24*)], and ii) non-thermal emission from electrons trapped in a magnetosphere (the surroundings of a celestial object where its magnetic field determines the motion of charged particles), forming a radiation belt (a magnetic axis with P.A.~ 0°).

In the first scenario, bipolar emission would produce different polarizations in each of the two components, because they would originate in different magnetic hemispheres (*25*). Alternatively, non-dipolar components could dominate the magnetic configuration of LSR J1835+3259, but we regard this as unlikely because the observed topologies of fully convective objects (*26*) indicate dipole-like arrangements for strong magnetic fields, which should include LSR J1835+3259 (*27*). There is no detectable circular polarization in the double-lobed structure in Figs. 3A and B, so we regard the bipolar emission hypothesis as improbable.

Non-thermal radio emission from emitting structures within the magnetosphere of LSR J1835+3259 is expected to produce a double morphology if a radiation belt is seen when the magnetic equator is aligned close to the line of sight, as it does for Jupiter (*28*). For LSR J1835+3259, this would need to occur at the rotation phase of the maps shown in Fig. 3, (magnetic axis on the plane of the sky). The location of the optical emission, and the absence of net circular polarization during these rotation phases, provide support for the presence of a radiation belt. In this scenario, we expect synchrotron radiation to dominate the detected radio emission, as it does for Jupiter (*28*). Such a configuration would produce radio emission with linear polarization (*28*); however, our observations were only sensitive to circular polarization.

## A radiation belt around LSR J1835+3259

A dipolar field constitutes a natural magnetic trap: the density of the magnetic field lines is lowest at the magnetic equator, and highest at the magnetic poles. In such a configuration, charged particles are forced to spiral along the magnetic field lines, bouncing back and forth between mirror points located in each magnetic hemisphere (*29*). A radiation belt around LSR J1835+3259 would therefore consist of high-energy, charged particles trapped in a dipolar magnetic field, similar to those present around the strongly magnetized planets of the Solar System (Earth, Jupiter, Saturn, Uranus and Neptune). It is not known how the particles are energized even in Jupiter's radiation belts (*30*). Whatever mechanism is acting, both a fast rotating central object and a strong magnetic field appear to be necessary to form and maintain a large and energetic radiation belt. Taking Jupiter as reference, electrons in the magnetosphere of LSR J1835+3259 would need to be accelerated up to tens of megaelectron-volts (MeV) (*28*). Assuming a magnetic field in a dipole-like configuration with a strength of 5 kG in the polar regions (*27*), the magnetic field strength in such a radiation belt would be ~2 G at a distance of 13 $R_\star$ from the optical photosphere (determined from burst B1; Fig. 3A). For our observing frequency of 5 GHz, this implies an average electron energy of 21 MeV (*21*). For B2, if the same radiation belt was also responsible for the extended structure visible in Fig. 3B, the average electron energy would need to have dropped to 8 MeV and the background magnetic field increased to 17 G. These estimates for the electron energy are similar to in situ measurements of Jupiter's radiation belts (*31*).

A radiation belt around LSR J1835+3259 with the typical energies estimated above would not be detectable during most of the object rotation, if the magnetic and rotation axes are misaligned. It is therefore necessary for Figs. 3A and B to be produced by a specific geometric configuration, in which the magnetic equatorial plane is seen edge-on and a beam of synchrotron emission is pointing towards Earth. We estimate that a misalignment angle (*β*) of a few degrees would make the radiation belt undetectable (less than $5\sigma$ at the observational sensitivity) in the maps before or after the bursts (*21*). In accordance with this prediction, we detect extended emission associated with the radiation belt only at rotation phases corresponding with the bursts (Fig. S1), which is the most favorable orientation.

The double-lobed structure is seen only once per rotation (Fig. 3 and Fig. S1); in our proposed radiation belt scenario this constrains the inclination angle (*i*) and its misalignment with the magnetic axis to be *i* = 130° and *β* = 40° (*21*) (Fig. 4C). There are no previous determinations of *i* for this UCD that can be used for comparison. However, we estimate *i* using a probabilistic inference method (*32*) using the rotation period (2.84140 ± 0.00039 hr), estimated radius (1.07 ± 0.05 $R_{Jup}$, *20*), and v sin(*i*) (43.9 ± 2.2 km s$^{-1}$, *33*). This yields a value of *i* = 120 ± 10°, similar to the value we estimated from the other method.

## Other sources of radio emission

Given the small beaming angle expected for synchrotron emission, the value we estimate for *β* implies that synchrotron emission from the radiation belt cannot be the only radio emission mechanism during the non-bursting phases. The maximum RCP emission does not coincide with the times where the radiation belt is detected (Fig. 2). Gyrosynchrotron emission and/or ECMI emission that is depolarized as it propagates through the magnetosphere (*19*) could alternatively be responsible for the non-bursting emission, in which case they would dominate emission at rotation phases other than those of B1 and B2 (Fig. S1).

The morphological differences between Fig. 3A and Fig. 3B could be due to physical and/or instrumental effects. Physical possibilities include time variability, and/or stretching of the magnetospheric field lines, as has been observed for Jupiter (*27, 34, 35*). Alternatively, the differing energetics of LSR J1835+3259 while bursting could indicate the presence of other radiation mechanisms in addition to synchrotron (such as gyrosynchrotron and/or depolarized ECMI emission). Instrumental effects could also cause the morphological differences between B1 and B2. Given the short duration of the bursts, and the cadence of observations of LSR J1830+3259 (*21*), the observations could have missed the exact rotation phase when the maximum of the LCP burst occurred.

The Stokes V maps of Fig. 3 show compact (milliarcsecond-scale) emission located near the optical position of LSR J1835+3259. We interpret this as auroral ECMI emission. Its detection at 5 GHz implies electron densities below $3 \times 10^{11}$ cm$^{-3}$ for the low density regions near the poles (*21*). We regard this density as plausible for LSR J1835+3259 (*2, 36*).

The radio powers of bursts B1 and B2 are two orders of magnitude higher than the total auroral power emitted by Jupiter (*21*), similar to previously reported bursts for this UCD (*3*). However, the profiles of B1 and B2 are substantially different (Fig. 2), which could be due to small changes in the physical parameters that determine the ECMI beam pattern (*37*). Burst B2 is 65% LCP polarized (B1 is 85%), is less intense, and has a slower decay. The energy and time decay are expected to be related: a slower decay indicates a longer trapping time of charged particles in the magnetosphere (*38*), which, in turn, depends on the energy of the electrons. According to this scenario, particles associated with the more energetic, fast-decaying B1 pulse quickly leave the magnetosphere, while those causing the less energetic burst B2 have a longer trapping time. The different profiles of B1 and B2 in the light curves (Fig. 2) could also affect the morphology of the radio emission (Figs. 3A and B); the longer decay time of burst B2 implies that ECMI emission was more affected by scattering and depolarization within the magnetosphere of LSR J1835+3259 than B1 was.

**Comparison with models**

We use previous radiation belt models (*10*) to interpret the morphology and light curve of both the bursting and non-bursting emission from LSR J1835+3259. Although the models were developed to explain the radio emission of much more massive stars (spectral types A or B), it has been proposed that they also apply to low-mass stars and gas giant planets (*10*). A diagram of our interpretation is shown in Fig. 4A. The model assumes a dipole-dominated magnetic field, with plasma permanently trapped by closed magnetic field lines, forming a radiation belt. The belt then produces, at least partially, the non-thermal radio emission (via the synchrotron mechanism, assuming similarity with Jupiter). Near the magnetic equator, electrons are accelerated by magnetic reconnection (the process where magnetic field lines rearrange and release stored energy) and propagate along the magnetic field lines, radiating non-thermal (gyrosynchrotron) emission. They eventually reach the poles of the central object, where they produce coherent pulsed ECMI auroral emission. In this interpretation, the ECMI emission is visible in the Stokes V map obtained during flares B1 and B2, whereas the corresponding Stokes I maps (non-bursting emission coincident in time with B1 and B2) show the morphology of the radiation belt. The belt is located close to the magnetic equator, at distances up to $\sim 13\,R_\star$ from the optical photosphere (Fig. 4B).

In the Jupiter system, the volcanic moon Io provides a source of plasma which supplies the radiation belt. It is possible that a satellite planet orbiting LSR J1835+3259 could play a similar role; numerous rocky exoplanets have been detected around M dwarf stars (e.g., *39, 40*). No companion to LSR J1835+3259 has been detected, but nor can one be ruled out (*21*). A planetary companion has previously been proposed to power the aurorae in this system (*3*). Its previously proposed parameters (*3*) would locate its orbit within the extended structure seen in our radio maps. We examined this possibility, but cannot either confirm or refute it (Supplementary Text). If such an exoplanet exists, it would be a potential source of plasma for the radiation belt.

**Acknowledgments:**

J.B.C. thanks A. Vidotto and J. Morin for useful conversations. We thank the anonymous referees for their constructive criticisms, which greatly improved the manuscript. The European VLBI Network is a joint facility of independent European, African, Asian, and North American radio astronomy institutes. This work has made use of data from the European Space Agency (ESA) mission Gaia, processed by the Gaia Data Processing and Analysis Consortium (DPAC). Funding for the DPAC has been provided by national institutions, in particular the institutions participating in the Gaia Multilateral Agreement.

**Funding:**

J.B.C., J.C.G. & J.M.M. were supported by Ministerio de Ciencia e Innovación/Agencia Estatal de Investigación grants PGC2018-098915-B-C22 and PID2020-117404GB-C22.

M.P.T. & L.P.M. were supported by Ministerio de Ciencia e Innovación with grants PID2020-117404GB-C21 and CEX2021-001131-S, and by funding from European Union NextGenerationEU grant PRTR-C17.I1.

J.B.C. & J.C.G. were supported by Generalitat Valenciana grants PROMETEO/2020-080, ASFAE/2022/018, and CIPROM/2022/64.

**Author contributions:**

Conceptualization: JBC, JCG.

Methodology: JBC, JCG, JMM, MPT.

Formal analysis: JBC, JCG.

Visualization: JBC, JCG, JMM, MPT, LPM.

Funding acquisition: JCG, JMM, MPT.

Writing – original draft: JBC, JCG.

Writing – review & editing: JCG, JBC, JMM, MPT, LPM.

**Competing interests:** The Authors declare that they have no competing interests.

**Data and materials availability:** The raw observations are available from the EVN data archive at http://archive.jive.nl/scripts/listarch.php under observation period 2021 and experiment ID EC077. The calibrated files are available at https://doi.org/10.5281/zenodo.8184165.

**Supplementary Materials**

Materials and Methods

Figs. S1 and S2

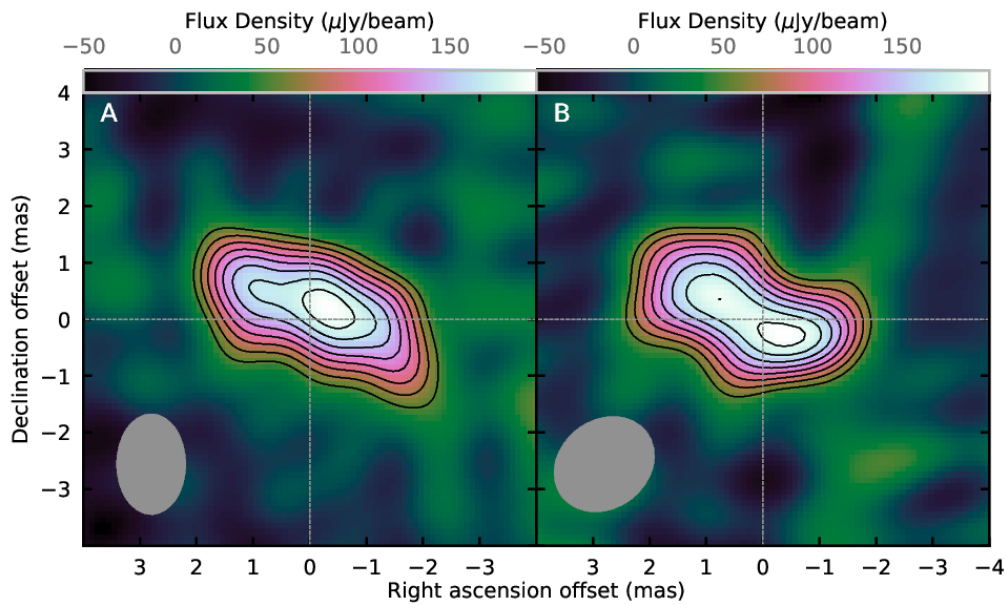

**Fig. 1. Reconstructed radio images of LSR J1835+3259 using the entire observations**. Stokes I reconstructed images of LSR J1835+3259 from the EVN observations on 2021 June 15. Contours indicate signal-to-noise ratios (S/N) of 3, 4, 5, 6, etc.; we required S/N ≥ 5 for a detection. (A) the first rotation and (B) second rotation of the object during the observations. The gray ellipses in the lower-left corners indicate the full-width at half-maximum (FWHM) beam sizes which have minor and major axes 1.14 × 1.70 mas at position angle 0.0°, and 1.49 × 1.81 mas at -53.9°, respectively. During each rotation the photosphere moved -0.36 mas in right

ascension and -0.19 mas in declination, which was corrected during data reduction (*21*). Both images are centered at the expected position of the optical photosphere (right ascension $18^h35^m37^s.75964$ and declination $32°59'37''.2595$; intersection of the grey lines), which was determined from the optical astrometry propagated to the observation epoch (*21*).

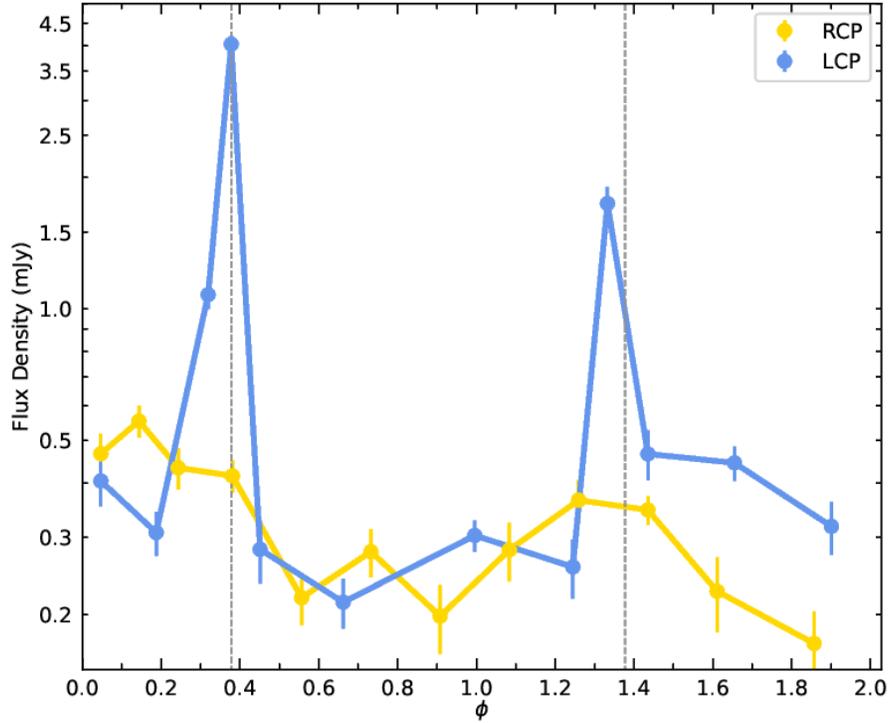

**Fig. 2. Light curves of the radio emission from LSR J1835 +3259 during the observations.** Data points are RCP (yellow) and LCP (blue) flux densities of LSR J1835+3259, determined by integrating the selected polarization flux density over the source region (*21*), plotted as a function of the rotation phase. The two vertical dashed lines are separated by the rotation period of the object and use the first burst as the reference phase. The vertical error bars indicate uncertainties in the flux densities at the $1\sigma$ level.

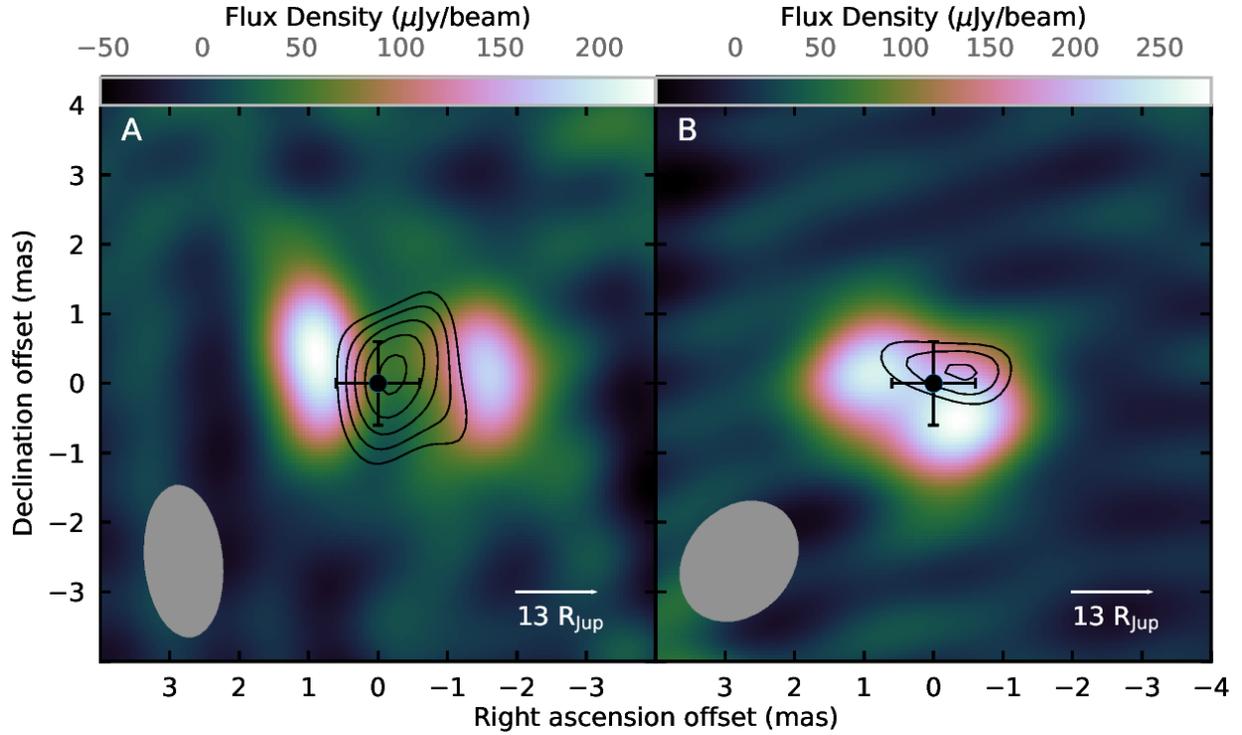

**Fig. 3. Reconstructed images of LSR J1835+3259 during the radio bursts.** Same as Figure 1, but for 30-minute windows around (A) burst B1 and (B) burst B2. 10 minutes of LCP data around each burst maximum were subtracted to produce the Stokes I images. The black contours show the Stokes V data for each of the 10 minute intervals around each burst, at S/N of 5, 6, 7, 8, and 9. The black circle indicates the expected size of the optical photosphere and its location derived from the optical astrometry, with error bars indicating the standard deviation resulting from our astrometric analysis (*21*).

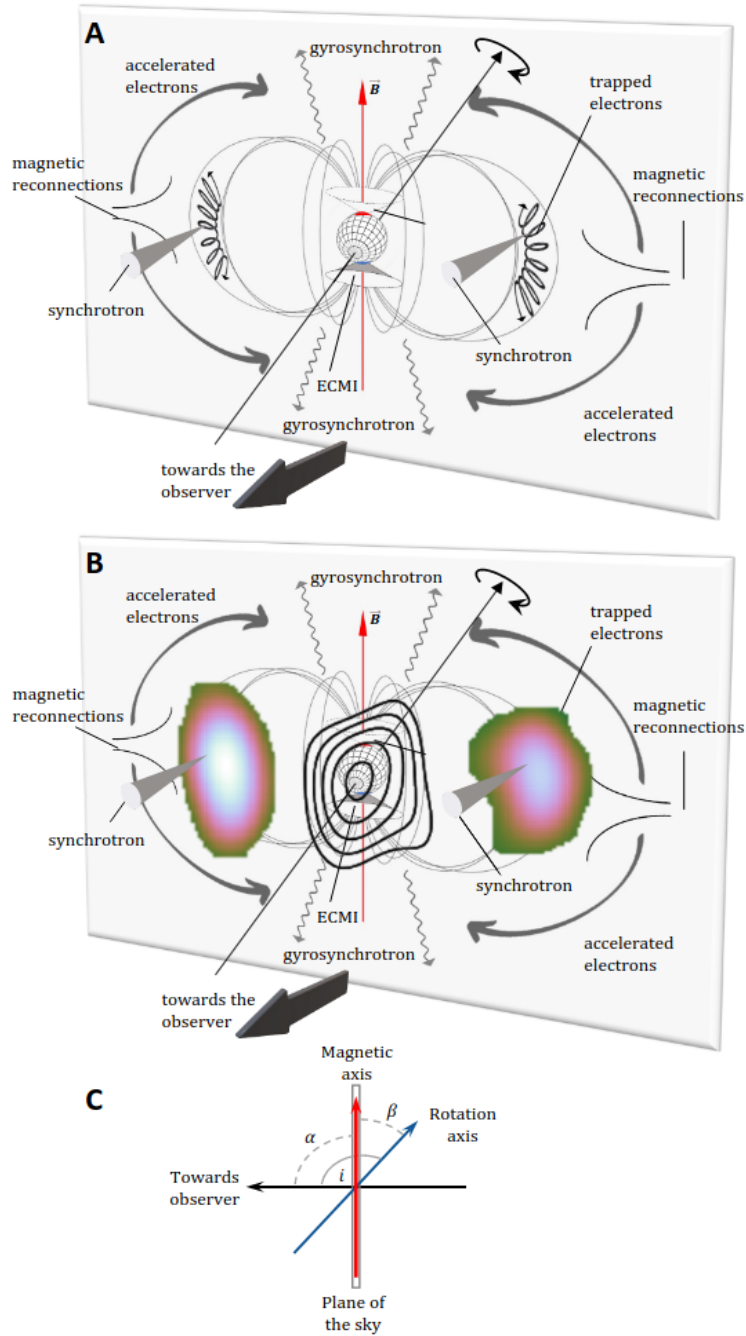

**Fig. 4. Diagram of the proposed magnetosphere configuration.** Our proposed magnetic configurations of the magnetosphere of LSR J1835+3259 at the time of the observations (not to scale). (A) Thin black lines around the central sphere (photosphere of LSR J1835+3259) represent the UCD's magnetic field lines; rotation and magnetic axes are indicated with black and red arrowed lines, respectively; hollow cones attached to the magnetic poles (red and blue caps) represent the ECMI, auroral emission; bold, curved arrows correspond to the trajectory of the accelerated electron towards the magnetic poles after magnetic reconnections, which generate gyrosynchrotron emission near the poles (wave arrows); helical lines at each side of the magnetosphere represent the trajectory of the trapped electrons near the equator, which produce

collimated synchrotron emission (narrow cones) originated at the radiation belts. (B) Same as panel A, but with the detected radio emission during burst B1 (Fig. 3A) superimposed. The separation between the different components of the radio emission have been deliberately enlarged with respect to those in Fig. 3A to better indicate their locations in the model. (C) Two-dimensional orientation of the magnetic and rotation axis during B1.

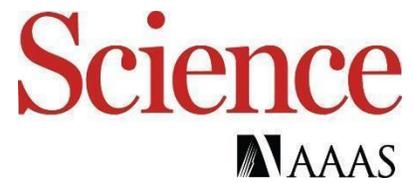

Supplementary Materials for

**Evidence for a radiation belt around a brown dwarf**

J. B. Climent, J. C. Guirado, M. Pérez-Torres, J. M. Marcaide, L. Peña-Moñino

Corresponding author: j.bautista.climent@uv.es

**The PDF file includes:**

Materials and Methods
Supplementary text
Figs. S1 to S2
References (42–50)

**Materials and Methods**

Data reduction and temporal analysis.

We observed LSR J1835+3259 with the EVN in 2021 with a recording aggregate bitrate of 4 Gbps, at a central frequency of 4.94 GHz and a total bandwidth of 256 MHz. A total of 14 antennas participated in the observations, which were performed in phase-referencing mode. We used B2 1846+32A, which is separated by 2.77° on the sky from the target, as a phase calibrator, with a duty cycle of 2.0 minutes on calibrator and 4.1 minutes on target. We reduced the data using the Astronomical Image Processing System (Aips, *42*), using standard routines. Given the high proper motion and parallax of LSR J1835+3259 (~0.4 mas per rotation), we applied time-based position corrections to the data, so the phase center tracked the same reference point of the object throughout the observations. We produced phase-referenced, channel-averaged images using the imaging program Difmap (*43*). We used the CLEAN algorithm within Difmap to produce images of the target in total flux and both circular polarizations. In each case, we centered a box at the maximum emission peak before using the task CLEAN. We iterated this procedure by adding boxes at the positions of successive maxima until no new peaks were distinguishable from the noise.

The rotationally averaged extended structures seen in Fig. 1 indicate that LSR J1835+3259 is spatially resolved (no resolved component was seen in the phase-reference calibrator; Fig. S2). The radio emission morphology of LSR J1835+3259 has similar orientation during two consecutive rotations (Fig. 1).

However, the morphology of the extended emission from LSR J1835+3259 varied within each rotation. This prevented us from performing a typical light-curve analysis, either by fitting Gaussian functions in the uv-plane or by calculating the discrete Fourier transform of the complex visibilities as a function of time. Instead, to produce the light-curve of the 2021 observations (Fig. 2), we split the dataset into subsets with the shortest time interval (adapting their length to the flux density and data quality) found to provide a 5σ detection in flux density. Given the limited uv-coverage of these snapshot maps, we compared with the interferometric beam in each interval to check that sidelobes were not present above the detection threshold in the recovered images. We repeated this procedure for LCP and RCP separately and used the beginning of our observations (20:10:51 UT on 2021 June 15) as the origin of the phases.

Optical and radio astrometry

We adopted optical astrometry from data release 3 of the Gaia mission (*16*). The Gaia DR3 catalog provides coordinates at epoch 2016.0 for LSR J1835+3259 (cataloged as Gaia DR3 2091177593123254016), which we propagated to the observing epoch using the Gaia DR3 values of proper motion and parallax. We find the optical coordinates of LSR J1835+3259 at epoch 2021 June 15th, 20:10:51 UT were right ascension $18^h35^m37^s.75964\pm0.00002$, declination $32°59'37''.2595\pm0.0002$ (J2000 equinox).

We compare these optical coordinates with the radio position of LSR J1835+3259 in our VLBI maps, after referencing the differential coordinates (resulting from our phase-reference analysis) to the International Celestial Reference Frame (ICRF) absolute coordinates of the phase-calibrator quasar B2 1846+32A. We performed an error analysis to determine the uncertainties of the radio coordinates due to shifts introduced by the propagation media, reference source structure, and the geometry of the interferometric array. These systematic contributions were five times larger than the limiting thermal noise uncertainty associated with the peak of brightness of the VLBI map. We determine a final astrometric error budget of 0.6 mas in each coordinate. The resulting comparison of the optical photosphere and the radio morphology of LSR J1835+3259, along with the corresponding uncertainties, is shown in Fig. 3.

The estimated position of the photosphere coincides, within the uncertainties, with the location of burst B1 (Fig. 3A). In an ECMI scenario, emission is expected at a frequency $\nu = s\nu_B$ where $s$ is the harmonic number (typically 1 or 2 for this mechanism) and $\nu_B = q_e B/(2\pi m_e)$ is the electron gyrofrequency, with $B$ being the magnetic field strength, and $q_e$ and $m_e$ the electron charge and mass, respectively. In the presence of a dipolar magnetic field, we can determine the magnetic field strength at any distance using $B = B_p r^{-3}$ where $B_p$ is the magnetic field strength at the poles. Taking the magnetic field strength measured on the surface of LSR J1835+3259 [5 kG (*27*)] as $B_p$, the ECMI mechanism at 5 GHz responsible for burst B1 would occur at a distance from the optical photosphere of 0.4 or 0.8 $R_\star$ for the first and second harmonic respectively.

Synchrotron emission from the radiation belt

We estimate the average electron energy in the radiation belt scenario using the expression

$$\frac{\nu_c}{MHz} = 16.08 \left(\frac{E}{MeV}\right)^2 \frac{B}{G} sin(\lambda) \qquad \text{(Equation S1)}$$

where $\nu_c$ is the critical frequency and $\lambda$ is the pitch angle (*44*). Above $\nu_c$, synchrotron emission falls off sharply. The maximum flux is emitted at a frequency of $0.29\nu_c$ (*28*). Assuming the maximum emission was at 5 GHz, and a pitch angle of 90°, we find E = 21 MeV for B = 2 G, and E = 8 MeV for B = 17 G.

Synchrotron radiation is emitted towards the direction of each electron's velocity, within a full-width half-maximum cone of $2/\gamma$, where $\gamma = 2$ E/MeV is the Lorentz factor. As such, a 21 MeV electron radiation cone would have an opening angle of ~ 3° (Fig. 4C). We use the simplification that this emission follows an isotropic distribution within such a cone. If the radiation belt in Fig. 3A is detected when the line of sight is parallel to the magnetic equatorial plane (maximum synchrotron emission), then a small tilt from this configuration of 1° would result in a signal below 5σ, assuming constant emission. This tilt would need to occur on time-scales shorter than the 30-min window employed to obtain the maps of Fig. 3, equivalent to less than 18% of the rotation phase. The misalignment between rotation and magnetic axes required to produce such wobble is > 3°. This is a small misalignment, compared to Solar system bodies (except for Saturn) and other celestial objects also described by an oblique dipole rotator like pulsars or magnetic chemically peculiar stars (*45*).

Geometry of the tilted dipole

Because the belt-like structures were seen once per rotation (Figs. 3A and B), the magnetic equatorial plane was also seen only once per rotation. Such a configuration can be produced when the inclination of the rotation axis ($i$) and its misalignment with the magnetic axis of the dipole ($\beta$) are related by $i \pm \beta = 90°$, where the plus and minus signs correspond to $i < 90°$ and $i > 90°$, respectively. At a rotation phase of $\phi \approx 0.17$ in our convention, the inclination of the magnetic axis with respect to the line of sight is $\alpha \sim 130°$ (*27*). At the same rotation phase, a similar value of $\alpha \sim 115°$ has previously been measured for this source (*46*). We therefore infer that from $\phi_1 \approx 0.17$ (previous $\alpha$ measurements) to $\phi_2 \approx 0.38$ (when ECMI occurs in our observations) the magnetic axis changes from an inclination of between 115° and 130° to 90°, because ECMI implies that the magnetic axis lies in the plane of sky. As the magnetic equator is seen only once per rotation, the corresponding 90° angle of the magnetic axis occurs only once per rotation. Therefore, 90° is the minimum inclination possible for the magnetic axis during a full rotation. This implies that $\beta = 0.25(\alpha-90°)/(\phi_2-\phi_1)$ is in the range of 30° to 50° and, consequently, $i$ lies between 120° and 140°, because $i - \beta = 90°$. We adopt the central values of these ranges, $\beta = 40°$ and $i = 130°$.

## Radio power of bursts

The flux densities of the bursting emission (3.7 ± 0.2 mJy and 1.5 ± 0.2 mJy) are equivalent to spectral luminosities of $1.4×10^{14}$ erg s$^{-1}$Hz$^{-1}$ and $5.8×10^{13}$ erg s$^{-1}$Hz$^{-1}$ for bursts B1 and B2, respectively. A visual inspection shows that the bursts are detected across the full 256 MHz bandwidth of our observations, so these values are lower limits due to emission outside the frequencies we are sensitive to. The implied radio power of bursts B1 and B2 are then >3.6 × $10^{22}$ erg s$^{-1}$ and >1.5 × $10^{22}$ erg s$^{-1}$ respectively, two orders of magnitude greater than the total auroral power emitted by Jupiter (*47*). The detection of ECMI emission at 5 GHz requires that plasma frequency (the frequency at which the electrons naturally oscillate) must be much lower than the electron gyrofrequency, implying electron densities below $3×10^{11}$ cm$^{-3}$ for the low density regions near the poles (*22*).

## Periodicity of the pulses

The time separation between B1 and B2 is close to, but different from, the previously measured rotation period of LSR J1835+3259. Assuming a rotating tilted dipole, pulse timing depends on the inclination ($i$), misalignment between the rotation and magnetic axes ($\beta$), and the deflection angle ($\theta$) of the radiation when refracted by plasma in the magnetosphere (*37, 48*). The deflection angle depends on the density of the plasma and any anisotropies in the medium (*37*). Changes in $\theta$ might then shift the pulse cadence with respect to the rotation period. If so, the time difference between pulses B1 and B2 in 2021 (~8.5 min; Fig. 2) could be due to a difference in the deflection angle of > 5°. We expect changes in the deflection angle between bursts B1 and B2 because the morphologies of the Stokes I maps during B1 and B2 (Figs. 3A and B) are also different, indicating that the magnetosphere properties vary.

**Supplementary text**

<u>Potential binary companion to LSR J1835+3259</u>

The double structure of radio emission from LSR J1835+3259 could be due to the presence of a very close companion body, either another brown dwarf or an exoplanet. If LSR J1835+3259 is a binary brown dwarf (mass ratio q ≤1), the morphology seen in our maps could be due to the emission of one or both objects and/or the interaction of their magnetospheres. No previous studies have proposed that LSR J1835+3259 is a binary brown dwarf. However, with the potential separation of the secondary object being ~2 mas (the separation in Fig. 1) such a companion would not have been visible in lower spatial resolution observations. Unresolved brown dwarf binaries have spectral differences compared to single objects of the same spectral type (*49*). Previous observations and modeling of the near-infrared spectrum of LSR J1835+3259 have determined the atmospheric parameters of this object (*31*). That study used the lines of NaI and TiO  and found no evidence of a companion object. Although this does not favor a binary companion to LSR J1835+3259, we cannot completely rule it out, given the broadening of the spectral lines produced by the rapid rotation of LSR J1835+3259 (which hampers any spectral analysis). Equal spectral-type binary brown dwarfs are particularly difficult to discern from their combined spectra (*49*). Other observations of this source using the transit method (*50*) have excluded Earth-sized transiting planets with orbital periods shorter than 1 day.

Our data constrain the properties of any binary companion. We assume that the size of the elongated east-west radio emission morphology corresponds to the semimajor axis of the possible binary orbit. Considering that the total mass of the system must be in the range 55 $M_{Jup}$ (q << 1) to 110 $M_{Jup}$ (q = 1), Kepler's third law limits the possible orbital periods to 30 to 50 hr. Such a fast orbital motion could affect our maps, given the time span of our observations, which would produce a change in the position angle (P.A.) of the structure of 15 to 30° for a face-on orbit. In the rotationally averaged maps of each period (Fig. 1), we do not detect any substantial changes in the P.A. of the structure. However, a small inclination of the companion orbit could cause the orbital motion to be undetectable in our maps. Therefore, we cannot confirm or refute the possibility that an unseen companion up to 55 $M_{Jup}$ might be orbiting LSR J1835+3259 at a very close distance.

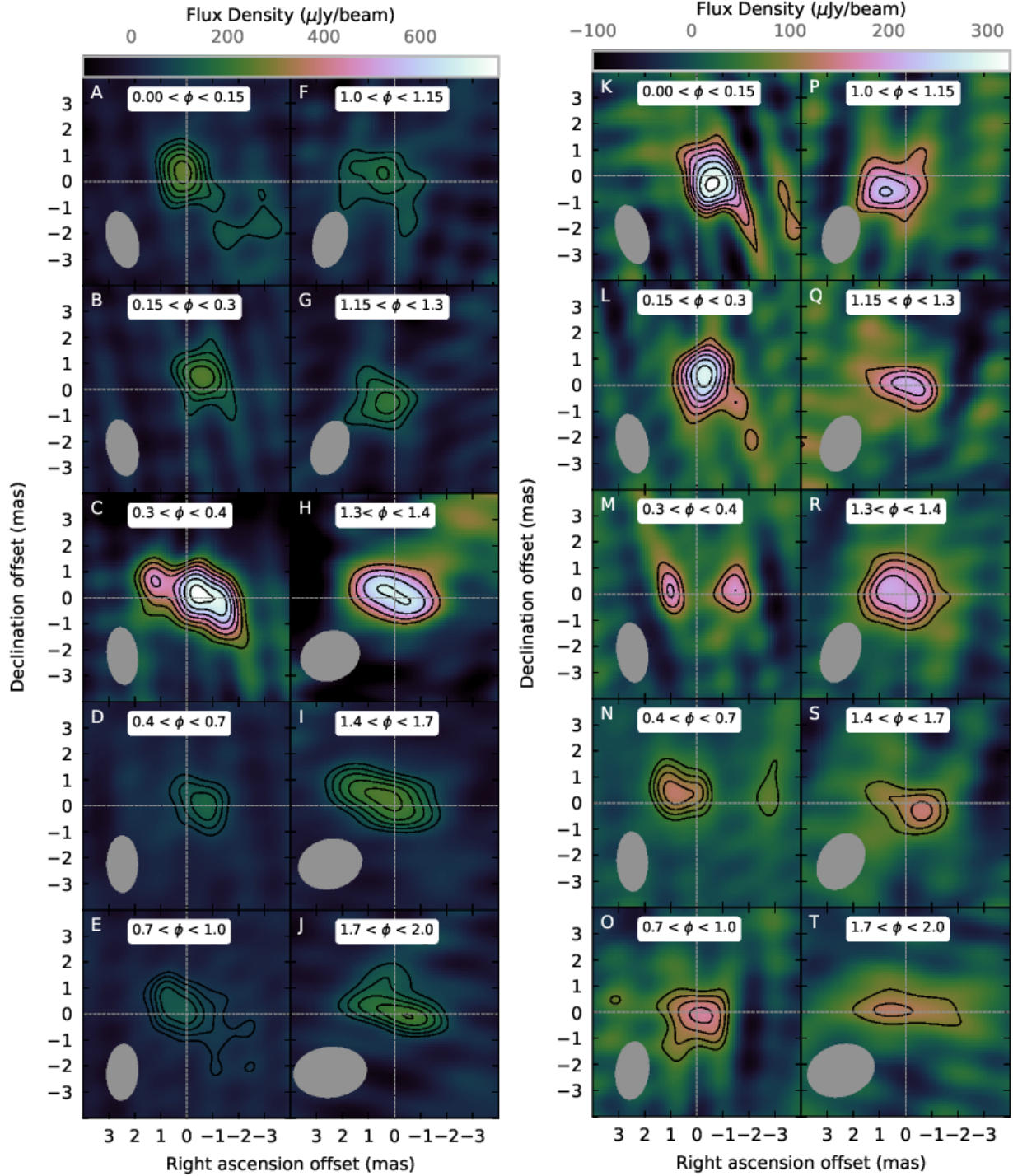

**Fig. S1. Temporal LCP and RCP images during the observations.** Same as Figure 1, but for the (A-J) LCP and (K-T) RCP images at different rotational phases. The phase range used for each map is labeled within the white boxes, with the phase origin ($\phi = 0$) at the beginning of our observations. Different color scales have been used for the LCP and RCP data, but are the same for all panels with the same polarization. The beam size (gray ellipse) increases during the observation as some antennas lose visibility of the source.

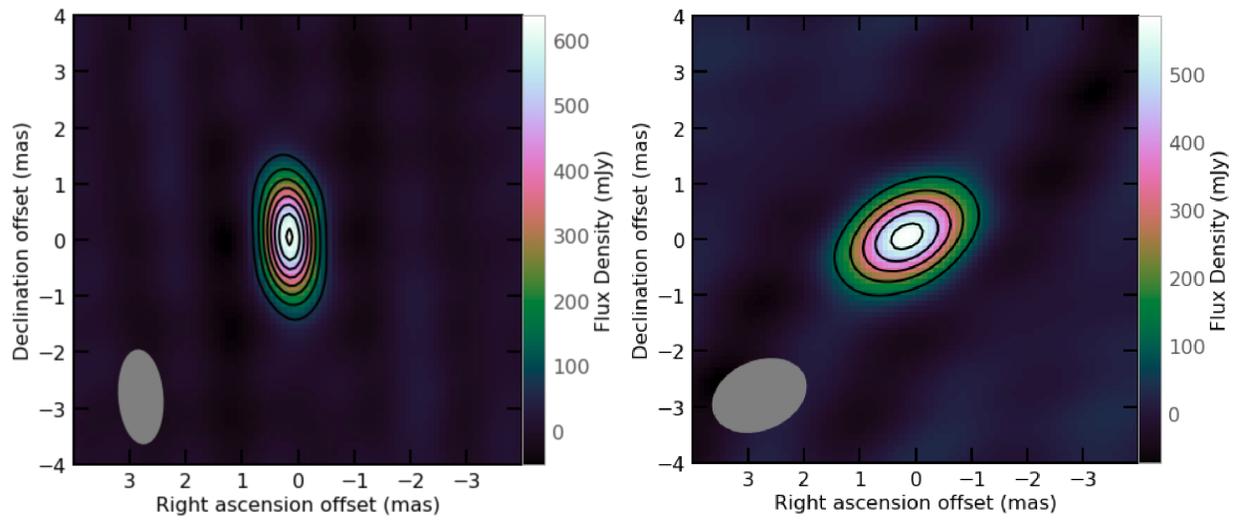

**Fig. S2. Reconstructed images of the phase calibrator.** The Stokes I reconstructed images of the phase calibrator B2 1846+32A are shown at the same time intervals as used in Fig. 3A and B. Contours indicate detection levels at 10σ, 20σ, 30σ, etc.